\def\ptitle{\tiny Closed-form sums $\dots$}
\font\tr=cmr12                          
\font\bf=cmbx12                         
\font\it=cmti12                         
\font\trbig=cmbx12 scaled 1500          
\font\tiny=cmr10                        
\output={\shipout\vbox{\makeheadline
                                      \ifnum\the\pageno>1 {\hrule}  \fi 
                                      {\pagebody}   
                                      \makefootline}
                   \advancepageno}

\headline{\noindent {\ifnum\the\pageno>1 
                                   {\tiny \ptitle\hfil
page~\the\pageno}\fi}}
\footline{}

\tr 
\def\ni{\noindent}             

\def\htab#1#2{{\hskip #1 in #2}}
\def\hi#1#2{$#1$\kern -2pt-#2} 
\def\hy#1#2{#1-\kern -2pt$#2$} 

\baselineskip 15 true pt  
\parskip=0pt plus 5pt 
\parindent 0.25in
\hsize 6.0 true in 
\hoffset 0.25 true in 
\emergencystretch=0.6 in                 
\vfuzz 0.4 in                            
\hfuzz  0.4 in                           
\vglue 0.1true in
\mathsurround=2pt                        
\topskip=24pt                            
\newcount\zz  \zz=0  
\newcount\q   
\newcount\qq    \qq=0  

\def\pref#1#2#3#4#5{\frenchspacing \global \advance \q by 1     
    \edef#1{\the\q}{\ifnum \zz=1{\item{$[{\the\q}]$}{#2}{\bf #3},{ #4.}{~#5}\medskip} \fi}}

\def\bref #1#2#3#4#5{\frenchspacing \global \advance \q by 1     
    \edef#1{\the\q}
    {\ifnum \zz=1 { %
       \item{$[{\the\q}]$} 
       {#2}, {\it #3} {(#4).}{~#5}\medskip} \fi}}

\def\gref #1#2{\frenchspacing \global \advance \q by 1  
    \edef#1{\the\q}
    {\ifnum \zz=1 { %
       \item{$^{\the\q}$} 
       {#2.}\medskip} \fi}}

 \def\sref #1{[#1]}

\def\references#1{\zz=#1
   \parskip=2pt plus 1pt   
   {\ifnum \zz=1 {\noindent \bf References \medskip} \fi} \q=\qq

\pref{\agub}{V. C. Aguilera-Navarro and R. Guardiola, J. Math.
Phys.}{32}{2135 (1991)}{}
\pref{\hald}{R. Hall, N. Saad and A. von Keviczky, J. Phys. A: Math. Gen.}{34}{1169 (2001)}{}
\pref{\hala}{R. Hall, N. Saad and A. von Keviczky, J. Math. Phys.}{39}{6345-51 (1998)}{}
\pref{\halb}{R. Hall and N. Saad, J. Phys. A: Math. Gen.}{33}{569 (2000)}{}
\pref{\halc}{R. Hall and N. Saad, J. Phys. A: Math. Gen.}{33}{5531 (2000)}{}
\bref{\luk1}{Yudell L. Luke}{The Special Functions and their Approximation}{Academic Press, 1969}{Page 116, formula 3}
\bref{\lukl}{Yudell L. Luke}{The Special Functions and their Approximation}{Academic Press, 1969}{Page 17, formula 5}
\pref{\ndlt}{Nota di Letterio Toscano, Boll. Un. Mat. Ital.}{3}{398 (1949)}{}
\bref{\luke}{Yudell L. Luke}{The Special Functions and their Approximation}{Academic Press, 1969}{Page 111, formula 40}
\bref{\grr}{G. E. Andrews, R. Askey and R. Roy}{Special Functions}{Cambridge University Press, 1999}{Theorem 2.2.5, Page 68}
\bref{\boch}{H. Buchholz}{The Confluent Hypergeometric Function}{Springer, 1969}{Page 144, last paragraph}
\bref{\elna}{Elna B. McBride}{Obtaining Generating Functions}{Springer, 1971}{Page 68, last paragraph}
}

\references{0}    

\htab{3.5}{CUQM-89}

\htab{3.5}{math-ph/0110042}

\htab{3.5}{October 2001}
\vskip 0.5 true in
\centerline{\bf\trbig Closed-form sums for some perturbation series involving }
\centerline{\bf\trbig associated Laguerre polynomials}
\medskip
\vskip 0.25 true in
\centerline{Richard L. Hall$^\dagger$, Nasser Saad$^\ddagger$ and Attila B. von Keviczky$^\dagger$}
\bigskip
{\leftskip=0pt plus 1fil
\rightskip=0pt plus 1fil\parfillskip=0pt\baselineskip 18 true pt
\obeylines
\baselineskip 10 pt
$^\dagger$Department of Mathematics and Statistics, Concordia University,
1455 de Maisonneuve Boulevard West, Montr\'eal, 
Qu\'ebec, Canada H3G 1M8.\par}

\medskip
{\leftskip=0pt plus 1fil
\rightskip=0pt plus 1fil\parfillskip=0pt\baselineskip 18 true pt
\obeylines
\baselineskip 10 pt
$^\ddagger$Department of Mathematics and Computer Science,
University of Prince Edward Island, 
550 University Avenue, Charlottetown, 
PEI, Canada C1A 4P3.\par}

\vskip 0.5 true in
\baselineskip 15 pt
\centerline{\bf Abstract}\medskip
Infinite series $\sum\limits_{n=1}^\infty {({\alpha\over 2})_n\over n}{1\over n!}{}_1F_1(-n,\gamma,x^2)$, where ${}_1F_1(-n,\gamma,x^2)={n!\over (\gamma)_n}L_n^{(\gamma-1)}(x^2),$
appear in the first-order perturbation correction for the wavefunction of the generalized spiked harmonic oscillator Hamiltonian $H=-{d^2\over dx^2}+B x^2 +{A\over x^2}+{\lambda\over x^\alpha}\quad 0\leq x<\infty,\quad \alpha,\lambda > 0,A\geq 0.$ It is proved that the series is convergent for all $x>0$ and $2\gamma>\alpha$ where $\gamma=1+{1\over 2}\sqrt{1+4A}$. Closed-form sums are presented for these series for the cases $\alpha=2, 4,$ and $6$. A general formula for finding the sum for  ${\alpha\over 2} = 2+m, m=0,1,2\dots$ in terms of associated Laguerre polynomials is also provided. 


\vfil\eject

\ni{\bf I. Introduction}

Aguillera-Navarro and Guardiola ${\sref{\agub}}$ encounter some difficulties inherent in connection with attempts to derive the first-order perturbation expansion to the wavefunction of the spiked harmonic oscillator Hamiltonian
$$
H=-{d^2\over dx^2}+ x^2 +{\lambda\over x^\alpha}\quad 0\leq x<\infty,\quad \alpha,\lambda > 0,\eqno(1.1)
$$
even for the case of $\alpha = 2$, where a complete exact solution is also available. The reason for these difficulties lies in computing infinite series of the type
$$\sum\limits_{n=1}^\infty {({\alpha\over 2})_n\over n}{1\over n!}{}_1F_1(-n;{3\over 2};x^2),\eqno(1.2)$$
where ${}_1F_1$ stands for the confluent hypergeometric function defined by
$${}_1F_1(-n;b;y)=\sum_{k=0}^{n}{(-n)_k\over (b)_k}{y^k\over k!}={n!\over (b)_n}L_n^{(b-1)}(y)\eqno(1.3)$$
in terms of the associated Laguerre polynomials $L_n^{(b-1)}(y),$ and $(a)_n,$ the shifted factorial (or {\it Pochhammer symbols}) defined by
$$(a)_0=1,\quad (a)_n=a(a+1)(a+2)\dots (a+n-1)={\Gamma(a+n)\over \Gamma(a)},\quad n=1,2,\dots.\eqno(1.4)$$

Recently, the present authors studied a more general Hamiltonian known now as the generalized spiked harmonic oscillator Hamiltonian ${\sref{\hald-\halc}}$
$$
H=H_0+\lambda V=-{d^2\over dx^2}+B x^2 +{A\over x^2}+{\lambda\over x^\alpha}\quad 0\leq x<\infty,\quad \alpha,\lambda > 0,A\geq 0,\eqno(1.5)
$$
defined on the one-dimensional space $(0\leq x<\infty)$ with eigenfunctions satisfying Dirichlet boundary conditions, that is to say, with wavefunctions vanishing at the boundaries. Herein Eq.(1.1) appears as a special case ($A=0, B=1$). They found that the matrix elements of the operator $x^{-\alpha}$, with respect to the exact solutions of the Gol'dman and Krivchenkov Hamiltonian $H_0$, namely
$$\psi_n(x)=(-1)^n\sqrt{{2B^{\gamma\over 2}\Gamma(n+\gamma)}\over n!\Gamma^2(\gamma)}x^{\gamma-{1\over 2}}e^{-{\sqrt{B}\over 2}x^2}{}_1F_1(-n,\gamma,\sqrt{B}x^2),\eqno(1.6)
$$
with exact eigenenergies
$$E_n=2\sqrt{B}(2n+\gamma),\quad n=0,1,2,\dots,\quad \gamma=1+{1\over 2}\sqrt{1+4A},\eqno(1.7)$$
are given explicitly by the following expressions:
$$
x_{mn}^{-\alpha}=(-1)^{n+m}B^{{\alpha\over 4}}{{({\alpha\over 2})_n}\over
(\gamma)_n}{{\Gamma(\gamma-{\alpha\over 2})}\over
\Gamma(\gamma)}\sqrt{{(\gamma)_n(\gamma)_m}\over {n!m!}}{}_3F_{2}(-m,\gamma-{\alpha\over
2},1-{\alpha\over 2};\gamma,1-n-{\alpha\over 2};1),\eqno(1.8)
$$
and valid for all values of the parameters  $\gamma$ and $\alpha$ such that $\alpha <2\gamma$. Furthermore, the matrix elements of the Hamiltonian Eq.(1.5) are given by
$$\eqalign{
H_{mn}=<m|H|n>\equiv 2\sqrt{B}(2n+\gamma)&\delta_{nm}+\lambda(-1)^{m+n}
B^{\alpha\over 4}
\sqrt{{(\gamma)_n(\gamma)_m\over n! m!}}
{{\Gamma(\gamma-{\alpha\over 2})({\alpha\over 2})_n}
\over (\gamma)_n\Gamma(\gamma)}\cr
&{}_3F_2(-m,\gamma-{\alpha\over 2},1-{\alpha\over 2};\gamma,1-{\alpha\over 2}-n;1).}\eqno(1.9)
$$
Of particular interest are the elements
$$
H_{0n}= \lambda(-1)^{n}
B^{\alpha\over 4}
\sqrt{(\gamma)_n\over n!}
{\Gamma(\gamma-{\alpha\over 2})\over \Gamma(\gamma)}{({\alpha\over 2})_n\over (\gamma)_n},\quad n\neq 0.\eqno(1.10)
$$
It is known that the first correction to the wavefunction by means of standard perturbation techniques leads to
$$\psi_0^{(1)}(x)=\sum\limits_{n=1}^{\infty}{H_{0n}\over {E_0-E_n}}\psi_n(x),\eqno(1.11)$$
where $H_{0n}$ and $\psi_n(x)$ are given by Eq.(1.10) and (1.6), respectively. Thus, the first correction to the wavefunction of the Hamiltonian Eq.(1.5) is given by
$$\psi_0^{(1)}(x)=-{B^{{\alpha\over 2}+{\gamma\over 4}-{1\over 2}}
\over 2\sqrt{2}}
{\Gamma(\gamma-{\alpha\over 2})\over \Gamma(\gamma)\sqrt{\Gamma(\gamma)}}
x^{\gamma-{1\over 2}}e^{-{\sqrt{B}\over 2}x^2}\sum\limits_{n=1}^\infty{({\alpha\over 2})_n\over n}{1\over n!}{}_1F_1(-n,\gamma,\sqrt{B}x^2).\eqno(1.12)
$$
The purpose of this article is to find closed-form sums for the infinite series appearing in Eq.(1.12), namely
$$\sum\limits_{n=1}^\infty {({\alpha\over 2})_n\over n}{1\over n!}{}_1F_1(-n;\gamma;x^2),\eqno(1.13)$$
where $2\gamma>\alpha$, $\alpha =2,4,6,\dots$ and we set $B=1,$ for simplicity. Because of Eq.(1.3), the results of this article can be expressed equally well in terms of the 
associated Laguerre polynomials. 
The importance of closed-form sums for the infinite series (1.13) is that they help us to understand the abnormal behavior of the standard, weak coupling, perturbation theory ${\sref{\agub}}$ for the singular Hamiltonians (1.1). 
Such infinite series were investigated earlier by the present authors $\sref{\hald}$, where they prove, in the case of $\alpha<2$, we have, by means of the inverse Laplace transform, that 
$$\sum\limits_{n=1}^\infty{({\alpha\over 2})_n\over n}{1\over n!}{}_1F_1(-n,\gamma,x^2)= {\Gamma(\gamma)\over 2\pi i} {\alpha\over 2} \int\limits_{c-i\infty}^{c+i\infty}e^{t}t^{-\gamma} (1-{x^2\over t}){}_3F_2(1+{\alpha\over 2},1,1;2,2;1-{x^2\over t})dt,\ c>0,\eqno(1.14)$$
where $|1-{x^2\over t}|<1$ which is indeed an important condition to insure the convergence of the series ${}_3F_2$ that appears on the right-hand side of (1.14). The functions  ${}_3F_{2}$ and ${}_1F_{1}$, mentioned above, are special cases of the generalized hypergeometric function 
$$
{}_pF_{q}(\alpha_1,\alpha_2,\dots,\alpha_p;\beta_1,\beta_2,\dots,\beta_q;z)=\sum\limits_{k=0}^\infty 
{\prod\limits_{i=1}^p(\alpha_i)_k\over  \prod\limits_{j=1}^q(\beta_j)_k}{z^k\over k!},\eqno(1.15)
$$ 
where $p$ and $q$ are non-negative integers and $\beta_j$ ($j=1,2,\dots,q$) none of which is equal to zero or to a negative integer. If the series does not terminate (either of $\alpha_i$, $i=1,2,\dots,p$, is negative integer), then the series converges or diverges according as $|z|<1$ or $|z|>1$. For $z=1$ on the other hand, the series is convergent, provided
$
{\sum\limits_{j=1}^q \beta_j-\sum\limits_{i=1}^p \alpha_i}>0.
$ 
This paper is organized as follows: in Sec. II we demonstrate that the infinite series on the left hand side of Eq.(1.14) converges for all $x>0$ and $\gamma>{\alpha\over 2}$. Furthermore, the integral representaion is still valid in such cases.
In Sec. III we prove that in the case $\alpha =2$, we have
$$\sum\limits_{n=1}^\infty {(1)_n\over n\ n!}{}_1F_1(-n;\gamma;x^2)=\psi(\gamma)-\log{x^2},\quad \gamma>1,$$
while in the case $\alpha =4$, we have
$$\sum\limits_{n=1}^\infty {(2)_n\over n\ n!}{}_1F_1(-n;\gamma;x^2)=\psi(\gamma)-\log{x^2}+{\gamma-1\over x^2}-1,\quad \gamma>2,$$
and for the case $\alpha =6$
$$\sum\limits_{n=1}^\infty {(3)_n\over n\ n!}{}_1F_1(-n;\gamma;x^2)=\psi(\gamma)-\log{x^2}+{\gamma-1\over x^2}-{3\over 2}+{(\gamma-1)(\gamma-2)\over 2x^4},\quad \gamma>3.$$
In Sec. IV we prove our main result that for ${\alpha\over 2}=2+m,m=0,1,2,\dots$
$$\eqalign{
\sum\limits_{n=1}^\infty {({\alpha\over 2})_n\over n\ n!}{}_1F_1(-n;\gamma;x^2)&=
\psi(\gamma)-\log x^2-(m+1)\sum\limits_{k=0}^m {(-m)_k\over (k+1)^2}\bigg(-{1\over x^2}\bigg)^k\bigg[L_k^{\gamma-1-k}(x^2)\cr
&
-{(\gamma-1)\over x^2}L_k^{\gamma-2-k}(x^2)\bigg]\cr}$$ where $L_n^{(a)}(\cdot)$ stands for the well-known associated Laguerre polynomials. An interpretation for the first-order correction of the wave function Eq.(1.12) as $x\rightarrow 0$ and some further remarks are given in Sec. V.
\medskip
\ni{\bf II. Integral Representation and the Convergence Problem}\medskip

In order to evaluate the sum in Eq.(1.13) for $\alpha >0$ and $2\gamma > \alpha$, we require a suitable integral representation of the confluent hypergeometric function ${}_1F_1(-n,\gamma, x^2)$ over an appropriate contour, in order to interchange summation with integration and thereby readily conclude the absolute convergence of the series just mentioned. We find the inverse Laplace transform (integral) representation ${\sref{\luk1}}$
$${}_1F_1(a,\gamma, x^2)={\Gamma(\gamma)\over 2\pi i}\int\limits_{c-i\infty}^{c+i\infty}e^{t}t^{-\gamma}(1-{x^2\over t})^{-a}dt\eqno(2.1)
$$
under the conditions $Re(\gamma)>0, c > 0,|\hbox{arg}(1-{x^2\over c})|< \pi$  (which is clearly true for $x$ real) to be most advantageous for achieving this end. 

Now turn to the evaluation of the summation in terms of the representation (2.1) written for $a=-n$, namely
$${}_1F_1(-n,\gamma, x^2)={\Gamma(\gamma)\over 2\pi i}\int\limits_{c-i\infty}^{c+i\infty}e^{t}t^{-\gamma}(1-{x^2\over t})^n dt,\quad\quad n=0,1,2,\dots,\eqno(2.2)$$
which substituted into the summation of Eq.(1.13) yields
$$\eqalign{
\sum\limits_{n=1}^\infty{({\alpha\over 2})_n\over n}{1\over n!}{}_1F_1(-n,\gamma,x^2)&=(2\pi i)^{-1}
\Gamma(\gamma)\sum\limits_{n=1}^\infty {({\alpha\over 2})_n\over n}{1\over n!}\int\limits_{c-i\infty}^{c+i\infty}e^{ t}t^{-\gamma}(1-{x^2\over t})^n dt\cr
&=(2\pi)^{-1}
\Gamma(\gamma)\sum\limits_{n=1}^\infty {({\alpha\over 2})_n\over n}{1\over n!}
\int\limits_{-\infty}^{\infty}e^{(c+iy)}(c+iy)^{-\gamma}
(1-{x^2\over c+iy})^n dy.}\eqno(2.3)
$$ 
The evaluation of this last infinite sum, involving integrations over the interval 
($-\infty,\infty$), is achieved by examining the summation of the integrand, namely
$$
\sum\limits_{n=1}^\infty {({\alpha\over 2})_n\over n}{1\over n!}
e^{ (c+iy)}(c+iy)^{-\gamma}(1-{x^2\over c+iy})^n
=e^{(c+iy)}(c+iy)^{-\gamma}\sum\limits_{n=1}^\infty {({\alpha\over 2})_n\over n}{1\over n!}(1-{x^2\over c+iy})^n,\eqno(2.4)
$$
and demonstrating that it has an $L_1(-\infty,\infty)$-majorant. Hence, the existence of such a majorant shall permit us to interchange summation with integration, as result of the Lebesgue Dominated Convergence Theorem. To arrive at such a majorant, we continue by noting that
$$\eqalign{
\sum\limits_{n=1}^\infty {({\alpha\over 2})_n\over n}{1\over n!}(1-{x^2\over c+iy})^n&=
\sum\limits_{n=0}^\infty {({\alpha\over 2})_{n+1}\over n+1}{1\over (n+1)!}(1-{x^2\over c+iy})^{n+1}\cr
&={\alpha\over 2}(1-{x^2\over c+iy})\sum\limits_{n=0}^\infty {({\alpha\over 2}+1)_{n}\over (2)_n}{(1)_n\over (2)_n}(1-{x^2\over c+iy})^{n}\cr
&={\alpha\over 2}(1-{x^2\over c+iy})\sum\limits_{n=0}^\infty {({\alpha\over 2}+1)_{n}(1)_n(1)_n\over (2)_n(2)_n}{(1-{x^2\over c+iy})^{n}\over n!}\cr
&={\alpha\over 2}(1-{x^2\over c+iy}){}_3F_2({\alpha\over 2}+1,1,1;2,2;1-{x^2\over c+iy})
\cr}\eqno(2.5)
$$
as consequence of $(a)_{n+1}=a(a+1)_n$, $n!=(1)_n$ and $(n+1)!=(2)_n$ . The series ${}_3F_2$, in Eq.(2.5), is convergent provided that $|1-{x^2\over c+iy}|<1$. Now, since 
$$1-{x^2\over c+iy}=1-{x^2(c-iy)\over c^2+y^2}=1-{x^2c\over c^2+y^2}+i{x^2y\over c^2+y^2},$$
for which
$$\bigg|1-{x^2\over c+iy}\bigg|^2=1-{x^2(2c-x^2)\over c^2+y^2}<1,$$
provided $c$ is chosen large enough - i. e. $x^2<2c$. For such $c$ we shall always have
$$0<1-{x^2(2c-x^2)\over c^2+y^2}<1\quad\quad \forall y\in R.\eqno(2.6)
$$ 
furthermore, the series ${}_3F_2$, in Eq.(2.5), is absolutely convergent for $|1-{x^2\over c+iy}|=1$, provided that $\alpha<2$ as a result of Eq.(1.15).
We now return to the majorization of summation Eq.(2.4), which entails
$$\eqalign{
\sum\limits_{n=1}^\infty
{
{({\alpha\over 2})_n}\over n\ n!}e^{(c+iy)}(c+iy)^{-\gamma}(1-{x^2\over c+iy})^n
&=
e^{c+iy}|c+iy|^{-\gamma}{\alpha\over 2}(1-{x^2\over c+iy}){}_3F_2({\alpha\over 2}+1,1,1;2,2;1-{x^2\over c+iy})
\cr
&<A(\alpha,c)|c+iy|^{-\gamma},}\eqno(2.7)
$$
where the convergence of ${}_3F_2$ and also $|1-{x^2\over c+iy}|<1$ were made use of. The most important aspect of inequality (2.7) is the appearance of the $L_1(-\infty,\infty)$-function $|c+iy|^{-\gamma}$ of variable $y$ majorizing the series
$$\sum\limits_{n=1}^\infty
{
{({\alpha\over 2})_n}\over n\ n!}|e^{\sqrt{B}(c+iy)}(c+iy)^{-\gamma}(1-{x^2\over c+iy})^n|,
$$
and this aspect justifies the evaluation of summation (2.4) by means of the Lebesgue Dominated Convergence Theorem. Thus we specifically have
$$\eqalign{
\sum\limits_{n=1}^\infty{({\alpha\over 2})_n\over n}{1\over n!}{}_1F_1(-n,\gamma,x^2)
&={\Gamma(\gamma)\over 
2\pi i}
\sum\limits_{n=1}^\infty {({\alpha\over 2})_n\over n}{1\over n!}
\int\limits_{-\infty}^{\infty}e^{(c+iy)}(c+iy)^{-\gamma}\times
(1-{x^2\over c+iy})^n idy\cr 
&={\Gamma(\gamma)\over 2\pi i}
\int\limits_{-\infty}^{\infty}e^{(c+iy)}(c+iy)^{-\gamma}
\bigg[\sum\limits_{n=1}^\infty {({\alpha\over 2})_n\over n}{1\over n!}(1-{x^2\over c+iy})^n\bigg]idy\cr
&={\Gamma(\gamma)\over 2\pi}
{\alpha\over 2}
\int\limits_{-\infty}^{\infty}e^{(c+iy)}(c+iy)^{-\gamma}
(1-{x^2\over c+iy})\times\cr
&\quad\quad\quad\quad\quad\quad{}_3F_2(1,1,1+{\alpha\over 2};2,2;1-{x^2\over c+iy})\ dy,\quad\quad\quad\quad\quad\quad (2.8)}
$$ 
which is an effective straight forward and precise determination of the summation $\sum\limits_{n=1}^\infty{({\alpha\over 2})_n\over n}{1\over n!}{}_1F_1(-n,\gamma,x^2)$ in terms of integrals of higher order hypergeometric function for arbitrary $\alpha < 2\gamma$. However, by utilizing $t=c+iy$ we reconvert the last expression of relation (2.8) to the inverse Laplace transform format, namely 
$$
\sum\limits_{n=1}^\infty{({\alpha\over 2})_n\over n}{1\over n!}{}_1F_1(-n,\gamma,x^2)=
{\Gamma(\gamma)\over 2\pi i}
{\alpha\over 2}
\int\limits_{c-i\infty}^{c+i\infty}e^{t}t^{-\gamma}
(1-{x^2\over t})\ {}_3F_2(1,1,1+{\alpha\over 2};2,2;1-{x^2\over t})\ dt\eqno(2.9)$$
valid for all $\alpha<2\gamma$. The computation of this expression is carried out in the next section.
\medskip
\ni{\bf III. Closed Form Sums}\medskip
\medskip
\noindent{\bf Lemma 1} For $\gamma>1$
$$
\sum\limits_{n=1}^\infty{1\over n}{}_1F_1(-n,\gamma,x^2)=
\psi(\gamma)-\log x^2\eqno(3.1)$$
\noindent{Proof:} For $\alpha =2$ and $(1)_n=n!$, Eq.(2.9) leads to
$$\eqalign{
\sum\limits_{n=1}^\infty{1\over n}{}_1F_1(-n,\gamma,x^2)&=
{\Gamma(\gamma)\over 2\pi i}
\int\limits_{c-i\infty}^{c+i\infty}e^{t}t^{-\gamma}
(1-{x^2\over t})\ {}_3F_2(2,1,1;2,2;1-{x^2\over t})dt\cr
&={\Gamma(\gamma)\over 2\pi i}
\int\limits_{c-i\infty}^{c+i\infty}e^{t}t^{-\gamma}
(1-{x^2\over t})\ {}_2F_1(1,1;2;1-{x^2\over t})dt\cr}
$$
It is known, however, that
$$
{}_2F_1(1,1;2;z)=-{1\over z}\log(1-z),\quad\quad\quad |z|<1$$
Thus, for $z=1-{x^2\over t}$, we have
$$\eqalign{
\sum\limits_{n=1}^\infty{1\over n}{}_1F_1(-n,\gamma,x^2)&=
-{\Gamma(\gamma)\over 2\pi i}
\int\limits_{c-i\infty}^{c+i\infty}e^{t}t^{-\gamma}
\log({x^2\over t})dt\cr
&=-\log x^2{\Gamma(\gamma)\over 2\pi i}
\int\limits_{c-i\infty}^{c+i\infty}e^{t}t^{-\gamma}\ dt+{\Gamma(\gamma)\over 2\pi i}
\int\limits_{c-i\infty}^{c+i\infty}e^{t}t^{-\gamma}
\log t\ dt\cr}
$$
The first integral on the right hand side can be computed by means of the reciprocal of the \hi{\Gamma}{function} ${\sref{\lukl}}$ or by means of the inverse Laplace transform of $f(t)=t^{-\gamma}$ for $\gamma>0$
$$[\Gamma(\gamma)]^{-1}= {1\over 2\pi i} \int\limits_{c-i\infty}^{c+i\infty}e^{t}t^{-\gamma}dt,\quad\quad c>0,\ \gamma>0 \eqno(3.2)$$
further, by differentiating Eq.(3.2) with respect to $\gamma$, we get
$${\psi(\gamma)\over \Gamma(\gamma)}={\Gamma^{\prime}(\gamma)\over [\Gamma(\gamma)]^2}={1\over 2\pi i}  \int\limits_{c-i\infty}^{c+i\infty}e^{t}t^{-\gamma}\log(t)\ dt,\quad\quad c>0,\ \gamma>0\eqno(3.3)$$
where $\psi(\gamma)$ is the digamma function defined as
$\psi(\gamma)={d\over d\gamma}\log\Gamma(\gamma)$. Therefore, 
$$\sum\limits_{n=1}^\infty{1\over n}{}_1F_1(-n,\gamma,x^2)=\psi(\gamma)-\log(x^2),\hbox{ for } \gamma>1$$
as required.\ \P

The result of Lemma 1 is not new indeed, and it was proved earlier by Toscano ${\sref{\ndlt}}$ by means of extensive used of calculus of finite difference. Toscano's result ${\sref{\ndlt}}$, however, was given in terms of associated Laguerre polynomials $L_n^{(\gamma)}(\cdot)$ where he proved that
$$
\sum\limits_{n=1}^\infty {(n-1)!\over \Gamma(n+\gamma)}L_n^{(\gamma-1)}(y)= {1\over \Gamma(\gamma)}[\psi(\gamma)-\log y]\eqno(3.4)
$$
For comparison, we use the relation between the confluent hypergeometric function ${}_1F_1(-n;\gamma+1;\cdot)$ and the associated Laguerre polynomials $L_n^{(\gamma)}(\cdot)$, namely
$$
{}_1F_1(-n,\gamma+1,\cdot)={\Gamma(n+1)\Gamma(\gamma+1)\over \Gamma(n+\gamma+1)}
L_n^{(\gamma)}(\cdot).\eqno(3.5)
$$ 
Thus, 
$$
\sum\limits_{n=1}^\infty
{1\over n}{}_1F_1(-n,\gamma,x^2)=\Gamma(\gamma)
\sum\limits_{n=1}^\infty {(n-1)!\over \Gamma(n+\gamma)}L_n^{(\gamma-1)}(x^2),\eqno(3.6)
$$
and this leads to the same results as lemma (1). In other words, Lemma (1) gives an independent proof of Toscano's result ${\sref{\ndlt}}$.

In order to find closed sums for Eq.(2.8) for positive even numbers of $\alpha$, we start with the reduction formula for ${}_3F_2(a,b,1;c,2;z)$ as given by Luke $\sref{\luke}$ 
$$z\ {}_3F_2(a,b,1;c,2;z)={(c-1)\over (a-1)(b-1)}\bigg[{}_2F_1(a-1,b-1;c-1;z)-1\bigg],\quad |z|<1.\eqno(3.7)$$
The purpose of the following lemma is to find the limit of Luke's identity as $b\rightarrow 1$.
\medskip
\noindent{\bf Lemma 2} For $a\neq 1$, $c\neq 1$, and $|{z\over z-1}|<1$,
$$z\ {}_3F_2(a,1,1;c,2;z)={(c-1)\over (a-1)}\bigg[{(c-a)\over (c-1)}\bigg({z\over z-1}\bigg){}_3F_2(c-a+1,1,1;c,2;{z\over z-1})-\log(1-z)\bigg]
,\eqno(3.8)$$
\noindent{Proof:}
From Pfaff's transformation ${\sref{\grr}}$ for ${}_2F_1$,
$$
{}_2F_1(a,b;c;z)=(1-z)^{-a}{}_2F_1(a,c-b;c;{z\over z-1}),\quad |{z\over z-1}|<1\eqno(3.9)
$$
which is also known as Euler's second identity, we have, by means of Eq.(2.1), that
$$z\ {}_3F_2(a,1,1;c,2;z)=\lim\limits_{b\rightarrow 1} {(c-1)\over (a-1)(b-1)}\bigg[(1-z)^{-(b-1)}{}_2F_1(c-a,b-1;c-1;{z\over z-1})-1\bigg]
$$
Using the identity
$$(1-z)^{-(b-1)}=e^{-(b-1)\log(1-z)},$$
and the series representation 
$$\eqalign{{}_2F_1(c-a,b-1;c-1;{z\over z-1})&=\sum\limits_{n=0}^\infty {(c-a)_n(b-1)_n\over (c-1)_n\ n!}\bigg({z\over z-1}\bigg)^{n}\cr
&=1+\sum\limits_{n=1}^\infty {(c-a)_n(b-1)_n\over (c-1)_n\ n!}\bigg({z\over z-1}\bigg)^{n}\cr}
$$
we have
$$\eqalign{
z\ {}_3F_2(a,1,1;c,2;z)=&\lim\limits_{b\rightarrow 1} {(c-1)\over (a-1)(b-1)}\bigg[
\bigg\{1-(b-1)\log(1-z)+{1\over 2}(b-1)^2[\log(1-z)]^2\cr
&
+O(b-1)^3\bigg\}\bigg\{1+\sum\limits_{n=1}^\infty {(c-a)_n(b-1)_n\over (c-1)_n\ n!}\bigg({z\over z-1}\bigg)^n\bigg\}-1\bigg].\cr}
$$
Further, since
$$\eqalign{
\sum\limits_{n=1}^\infty {(c-a)_n(b-1)_n\over (c-1)_n n!}\bigg({z\over z-1}\bigg)^n&=\sum\limits_{n=0}^\infty {(c-a)_{n+1}(b-1)_{n+1}\over (c-1)_{n+1} (n+1)!}\bigg({z\over z-1}\bigg)^{n+1}\cr
&={(c-a)(b-1)\over (c-1)}\bigg({z\over z-1}\bigg)
\sum\limits_{n=0}^\infty {(c-a+1)_{n}(b)_{n}(1)_n\over (c)_n (2)_n n!}\bigg({z\over z-1}\bigg)^{n}\cr
&={(c-a)(b-1)\over (c-1)}\bigg({z\over z-1}\bigg){}_3F_2(c-a+1,b,1;c,2;{z\over z-1}),
\cr}
$$
where we implement $(a)_{n+1}=a(a+1)_n$ and the series representation of ${}_3F_2$ by means of Eq.(1.15), thus we have
$$\eqalign{
z\ {}_3F_2(a,1,1;c,2;z)&=\lim\limits_{b\rightarrow 1} {(c-1)\over (a-1)(b-1)}\bigg[
{(c-a)(b-1)\over (c-1)}\bigg({z\over z-1}\bigg){}_3F_2(c-a+1,b,1;c,2;{z\over z-1})-\cr
&\quad\quad(b-1)\log(1-z)-{(c-a)(b-1)^2\over (c-1)}\bigg({z\over z-1}\bigg)\log(1-z)\times\cr
&\quad\quad{}_3F_2(2-{\alpha\over 2},b,1;2,2;{z\over z-1})+O(b-1)^2\bigg]\cr
&={(c-1)\over (a-1)}\bigg[{(c-a)\over (c-a)}\bigg({z\over z-1}\bigg){}_3F_2(c-a+1,1,1;c,2;{z\over z-1})-\log(1-z)\bigg]\cr}
$$
this proves the lemma\ \P.

As a direct application of this lemma, we have for $a=1+{\alpha\over 2},$ and $c=2$, 
$$z\ {}_3F_2(1+{\alpha\over 2},1,1;2,2;z)={2\over \alpha}\bigg[(1-{\alpha\over 2}){z\over z-1}{}_3F_2(2-{\alpha\over 2},1,1;2,2;{z\over z-1})-\log(1-z)\bigg],\quad \bigg|{z\over z-1}\bigg|<1 \eqno(3.10)$$
For the purpose of our applications, where we have $z=1-{x^2\over t}$ for $t=c+iy$, we must note 
$$|{z\over z-1}|^2=({z\over z-1})({\overline{z}\over {\overline{z}-1}})<1$$
which leads to $\Re(z)<{1\over 2}$. However, the real part of $z=1-{x^2\over c+iy}=1-{x^2(c-iy)\over c^2+y^2}$ is
$$1-{x^2c\over c^2+y^2}<{1\over 2}\rightarrow {1\over 2}<{x^2c\over c^2+y^2}<{x^2\over c}$$
that to say ${c\over 2}< x^2$ which does not contradict our requirement as given by Eq.(2.5). Therefore, Lemma 2 can be used with arbitrary values of $\alpha$ provided that $\alpha<2\gamma$. However, for $2-{\alpha\over 2}=-m,\quad m=0,1,2,\dots$, the series ${}_3F_2$ on the right hand side of Eq.(3.8) terminates and the convergence problem does not arise.
 In this case we have 
\medskip
\noindent{\bf Lemma 3} For $2-{\alpha\over 2}=-m,\ m=0,1,2,\dots$, we have
$$\eqalign{ \sum\limits_{n=1}^\infty{({\alpha\over 2})_n\over n}&{1\over n!}{}_1F_1(-n,\gamma,x^2)= \cr &(1-{\alpha\over 2}){\Gamma(\gamma)\over 2\pi i} \int\limits_{c-i\infty}^{c+i\infty}e^{t}t^{-\gamma} (1-{t\over x^2}){}_3F_2(2-{\alpha\over 2},1,1;2,2;1-{t\over x^2})\ dt+\psi(\gamma)-\log x^2}\eqno(3.11)$$

\noindent{Proof:} Using Lemma 2, and the fact that $z= 1 - {x^2\over t}$, Eq. (2.9) leads to
$$\eqalign{\sum\limits_{n=1}^\infty{({\alpha\over 2})_n\over n}{1\over n!}{}_1F_1(-n,\gamma,x^2)=&
(1-{\alpha\over 2}){\Gamma(\gamma)\over 2\pi i} \int\limits_{c-i\infty}^{c+i\infty}e^{t}t^{-\gamma} (1-{t\over x^2}){}_3F_2(2-{\alpha\over 2},1,1;2,2;1-{t\over x^2})\ dt\cr
&-{\Gamma(\gamma)\over 2\pi i} \int\limits_{c-i\infty}^{c+i\infty}e^{t}t^{-\gamma}\log({x^2\over t})\ dt}\eqno(3.12)$$
The second integral on the right hand side of Eq.(3.12) is already computed by means of Lemma 1 and leads to
$${\Gamma(\gamma)\over 2\pi i} \int\limits_{c-i\infty}^{c+i\infty}e^{t}t^{-\gamma}\log({x^2\over t})\ dt=\log x^2-\psi(\gamma)$$
which completes the proof of the lemma\ \P.
\medskip
\noindent{\bf 3.1. The case $\alpha=4$}\medskip 
In this case $2-{\alpha\over 2}=0$, and Lemma 3 leads to
$$\eqalign{\sum\limits_{n=1}^\infty{(2)_n\over n\ n!}{}_1F_1(-n,\gamma,x^2)&=-
{\Gamma(\gamma)\over 2\pi i}  \int\limits_{c-i\infty}^{c+i\infty}e^{t}t^{-\gamma}
(1-{t\over x^2})\ dt +\psi(\gamma)-\log x^2\cr
&=-{\Gamma(\gamma)\over 2\pi i} \int\limits_{c-i\infty}^{c+i\infty}e^{t}t^{-\gamma}
dt+{1\over x^2}{\Gamma(\gamma)\over 2\pi i}\int\limits_{c-i\infty}^{c+i\infty}e^{t}t^{-\gamma+1}
dt\cr
&\quad\quad\quad+\psi(\gamma)-\log x^2\cr
&={\gamma-1\over x^2}-1+\psi(\gamma)-\log{x^2},\quad \gamma>2,
\cr}\eqno(3.1.1)$$
where we invoke Eq.(3.2). There is, indeed, an independent confirmation for this result. 
Since $(2)_n = (1+n)(1)_n$, the infinite sum in Eq.(1.13), reads
$$\sum\limits_{n=1}^\infty {1+n\over n}{}_1F_1(-n;{\gamma};x^2)= \sum\limits_{n=1}^\infty {1\over n}{}_1F_1(-n;{\gamma};x^2)+\sum\limits_{n=1}^\infty {}_1F_1(-n;{\gamma};x^2),\eqno(3.1.2)$$
The first series on the right hand side is summable by means of Toscano's result ${\sref{\ndlt}}$ (regardless the integral representation). For the second sum on the right hand side, we refer to Buchholz's identity ${\sref{\boch}}$, 
$$\sum_{n=0}^\infty {(-\nu)_n\Gamma(\gamma+\nu+1)\over \Gamma(n+\gamma+1)}L_n^{(\gamma)}(y)=y^\nu,\quad \gamma+\nu>-1,\ \nu\neq 0,1,2,\dots\eqno(3.1.3)$$ 
using Eq.(3.5), we have
$$\sum_{n=0}^\infty {(-\nu)_n\Gamma(\gamma+\nu+1)\over n! \Gamma(\gamma+1)}{}_1F_1(-n;\gamma+1;y)=y^\nu,\quad \gamma+\nu>-1\eqno(3.1.4)$$
Setting $\nu=-1$, we get
$$\sum_{n=0}^\infty {}_1F_1(-n;\gamma+1;y)=\gamma y^{-1},\quad \gamma>0$$
or
$$\sum_{n=1}^\infty {}_1F_1(-n;\gamma;x^2)={\gamma-1\over x^2}-1,\quad \gamma>1\eqno(3.1.5)$$
and thus 
$$\sum\limits_{n=1}^\infty{({2})_n\over n}{1\over n!}{}_1F_1(-n;\gamma;x^2)=\psi(\gamma)-\log x^2+{\gamma-1\over x^2}-1$$
which confirm our result as given by Eq.(3.1.1).
\medskip
\noindent{\bf 3.2. The case $\alpha=6$}
\medskip
By means of Eq.(3.10), we have 
$$\eqalign{\sum\limits_{n=1}^\infty{(3)_n\over n\ n!}{}_1F_1(-n,\gamma,x^2)&=-2{\Gamma(\gamma)\over 2\pi i}  \int\limits_{c-i\infty}^{c+i\infty}e^{t}t^{-\gamma}(1-
{t\over x^2})(1-{1\over 4}(1-{t\over x^2}))\ dt+\psi(\gamma)-\log x^2\cr
&={\Gamma(\gamma)\over 2\pi i}  \int\limits_{c-i\infty}^{c+i\infty}e^{t}t^{-\gamma}\bigg(
-{3\over 2}+{t\over x^2}+{t^2\over 2x^4}\bigg)\ dt+\psi(\gamma)-\log x^2\cr
&=-{3\over 2}{\Gamma(\gamma)\over 2\pi i}  \int\limits_{c-i\infty}^{c+i\infty}e^{t}t^{-\gamma}\ dt+{1\over x^2}{\Gamma(\gamma)\over 2\pi i}  \int\limits_{c-i\infty}^{c+i\infty}e^{t}t^{-\gamma+1}\ dt \cr
& +{1\over 2x^4}{\Gamma(\gamma)\over 2\pi i}  \int\limits_{c-i\infty}^{c+i\infty}e^{t}t^{-\gamma+2}\ dt +\psi(\gamma)-\log x^2\cr
&=-{3\over 2}+{1\over x^2}
{\Gamma(\gamma)\over \Gamma(\gamma-1)}+{1\over 2x^4}
{\Gamma(\gamma)\over \Gamma(\gamma-2)}+\psi(\gamma)-\log x^2\cr
&=-{3\over 2}+{\gamma-1\over x^2}+{(\gamma-1)(\gamma-2)\over 2x^4}+\psi(\gamma)-\log x^2,\quad \gamma>3,\quad (3.2.1)
\cr}$$
where we invoke Eq.(3.2). These results can be also confirmed by an independent proof. Since $(3)_n={1\over 2} (n^2+3n+2) (1)_n,$ the infinite series Eq.(1.13) becomes in this case
$$\sum\limits_{n=1}^\infty {({3})_n\over n}{1\over n!}{}_1F_1(-n;\gamma;x^2)={1\over 2}\sum\limits_{n=1}^\infty n\ {}_1F_1(-n;\gamma;x^2)+{3\over 2}\sum\limits_{n=1}^\infty  {}_1F_1(-n;\gamma;x^2)+\sum\limits_{n=1}^\infty{1\over n} {}_1F_1(-n;\gamma;x^2)$$
The second and third series on the right hand side are summable by means of Eq.(3.1.5) and Eq.(3.6) respectively, regardless the integral representation. For the first series on the right hand side, it is enough to take $\nu=-2$ in Eq.(3.1.4) to conclude that 
$$\sum\limits_{n=1}^\infty n\ {}_1F_1(-n;\gamma;x^2)={(\gamma-1)(\gamma-2)\over x^4}-{\gamma-1\over x^2},\quad \gamma>2\eqno(3.2.2)
$$
This leads to
$$\sum\limits_{n=1}^\infty {({3})_n\over n}{1\over n!}{}_1F_1(-n;\gamma;x^2)=
{\gamma-1\over x^2}-{3\over 2}+{(\gamma-1)(\gamma-2)\over 2x^4}-\log{x^2}+\psi(\gamma),\quad \gamma>3
$$
\noindent{\bf IV. General Case}
\medskip

The results just mentioned for $\alpha=4$ and $\alpha=6$ can be  generalized indeed to any $\alpha$ such that $2-{\alpha\over 2}=-m,m=0,1,2,\dots$. 
\medskip
\noindent{\bf Lemma 4} For $2-{\alpha\over 2}=-m,\ m=0,1,2,\dots$ and $\gamma>{\alpha\over 2}$, we have
$$\eqalign{
\sum\limits_{n=1}^\infty {({\alpha\over 2})_n\over n\ n!}{}_1F_1(-n;\gamma;x^2)=&
\psi(\gamma)-\log x^2-(m+1)\sum\limits_{k=0}^m {(-m)_k(1)_k\over (2)_k(2)_k}\bigg[{}_2F_0(-k,1-\gamma;-;-{1\over x^2})\cr
&-{\gamma-1\over x^2}{}_2F_0(-k,2-\gamma;-;-{1\over x^2})\bigg].\quad\quad\quad\quad\quad\quad\quad\quad\quad\quad\quad\quad(4.1)\cr}$$
\noindent{Proof:}
Using Lemma 3, we have for $2-{\alpha\over 2}=-m,\ m=0,1,2,\dots$, 
$$\sum\limits_{n=1}^\infty{({\alpha\over 2})_n\over n}{1\over n!}{}_1F_1(-n,\gamma,x^2)= \psi(\gamma)-\log x^2-(m+1){\Gamma(\gamma)\over 2\pi i} \int\limits_{c-i\infty}^{c+i\infty}e^{t}t^{-\gamma} (1-{t\over x^2}){}_3F_2(-m,
1,1;2,2;1-{t\over x^2})\ dt$$
The function ${}_3F_2(-m,
1,1;2,2;1-{t\over x^2})$ is a terminated series, specifically a polynomial of degree $m$, and therefore we may integrate term by term using the series representation of ${}_3F_2.$ We have
$$\eqalign{I_m^\gamma(x)&=-(m+1){\Gamma(\gamma)\over 2\pi i} \int\limits_{c-i\infty}^{c+i\infty}e^{t}t^{-\gamma} (1-{t\over x^2}){}_3F_2(-m,
1,1;2,2;1-{t\over x^2})\ dt\cr
&=-(m+1){\Gamma(\gamma)\over 2\pi i} \sum\limits_{k=0}^m{(-m)_k(1)_k\over (2)_k(2)_k}\bigg[\int\limits_{c-i\infty}^{c+i\infty}e^tt^{-\gamma}(1-{t\over x^2})^k\ dt-{1\over x^2}\int\limits_{c-i\infty}^{c+i\infty}e^tt^{-\gamma+1}(1-{t\over x^2})^k\ dt\bigg].\cr}
$$
Since
$$(1-{t\over x^2})^k=\sum\limits_{l=0}^k {(-k)_l\over l!}({t\over x^2})^l,\hbox{ finite number of terms},
$$
we have
$$\eqalign{
I_m^\gamma(x)&=-(m+1) \sum\limits_{k=0}^m{(-m)_k(1)_k\over (2)_k(2)_k}\bigg[
\sum\limits_{l=0}^k{(-k)_l\over l!}{1\over x^{2l}}{\Gamma(\gamma)\over 2\pi i}\int\limits_{c-i\infty}^{c+i\infty}e^tt^{-\gamma+l}\ dt\cr
&\quad\quad -\sum\limits_{l=0}^k{(-k)_l\over l!}{1\over x^{2l+2}}{\Gamma(\gamma)\over 2\pi i}\int\limits_{c-i\infty}^{c+i\infty}e^tt^{-\gamma+l+1}\ dt\bigg]\cr
&=-(m+1) \sum\limits_{k=0}^m{(-m)_k(1)_k\over (2)_k(2)_k}\bigg[
\sum\limits_{l=0}^k{(-k)_l\over l!}{1\over x^{2l}}{\Gamma(\gamma)\over \Gamma(\gamma-l)}-\sum\limits_{l=0}^k{(-k)_l\over l!}{1\over x^{2l+2}}{\Gamma(\gamma)\over \Gamma(\gamma-l-1)}\bigg]
\cr}
$$
where we have used Eq.(3.2) for $\gamma>l+1$. From the identity $\Gamma(\gamma-l)=\Gamma(\gamma)(\gamma)_{-l}=\Gamma(\gamma){(-1)^l\over (1-\gamma)_l}$ and $\Gamma(\gamma-l-1)=\Gamma(\gamma-1)(\gamma-1)_l=\Gamma(\gamma-1){(-1)^l\over (2-\gamma)_l}$, we have now
$$I_m^\gamma(x)=-(m+1) \sum\limits_{k=0}^m{(-m)_k(1)_k\over (2)_k(2)_k}\bigg[
\sum\limits_{l=0}^k{(-k)_l(1-\gamma)_l\over l!}(-{1\over x^{2}})^l-{(\gamma-1)\over x^2}\sum\limits_{l=0}^k{(-k)_l(2-\gamma)_l\over l!}(-{1\over x^{2}})^l\bigg]$$
which finally leads to
$$I_m^\gamma(x)=-(m+1)\sum\limits_{k=0}^m {(-m)_k(1)_k\over (2)_k(2)_k}\bigg[{}_2F_0(-k,1-\gamma;-;-{1\over x^2})-{(\gamma-1)\over x^2}{}_2F_0(-k,2-\gamma;-;-{1\over x^2})\bigg]$$
by means of Eq.(1.15)\ \P.

The significance of this lemma is that the infinite series of Eq.(2.9) can now be replaced by a finite series that is much easier to calculate. To illustrate the use of this lemma, we shall find now the infinite series of Eq.(2.9) for the case $\alpha=8$, i.e. $m=2$, since
$$\eqalign{
\sum\limits_{n=1}^\infty {({4})_n\over n\ n!}{}_1F_1(-n;\gamma;x^2)=&
\psi(\gamma)-\log x^2-3\sum\limits_{k=0}^2 {(-m)_k(1)_k\over (2)_k(2)_k}\bigg[{}_2F_0(-k,1-\gamma;-;-{1\over x^2})\cr
&-{\gamma-1\over x^2}{}_2F_0(-k,2-\gamma;-;-{1\over x^2})\bigg]\cr}$$
and since
$${}_2F_0(0,1-\gamma;-;-{1\over x^2})=1$$ 
$${}_2F_0(-1,1-\gamma;-;-{1\over x^2})=1-{(\gamma-1)\over x^2}$$
$${}_2F_0(-2,1-\gamma;-;-{1\over x^2})=1-{2(\gamma-1)\over x^2}+{(\gamma-1)(\gamma-2)\over x^4}$$
and similarly for ${}_2F_0(-k,2-\gamma;-;-{1\over x^2})$, $k=0,1,2$. It is straightforward calculation to find a closed-form sum for the infinite series Eq.(2.9) for $\alpha =8$ which leads in this case to
$$\sum\limits_{n=1}^\infty {(4)_n\over n\ n!}{}_1F_1(-n;\gamma;x^2)=\psi(\gamma)-\log{x^2}-{11\over 6}+{(\gamma-1)(\gamma-2)(\gamma-3)\over 3x^6}+{(\gamma-1)(\gamma-2)\over 2x^4}+{(\gamma-1)\over x^2},\eqno(4.2)$$ 
valid for $\gamma>4$. It is interesting to mention here that the result of Lemma 4, can be written in terms of the well-known associated Laguerre polynomials. Indeed, from the identity ${\sref{\elna}}$
$$(-1)^n L_k^{a-n}(y)= {y^n\over n!}{}_2F_0(-n,-a;-;-{1\over y}),$$
the result of Lemma 4 can be written as
$$\eqalign{
\sum\limits_{n=1}^\infty {({\alpha\over 2})_n\over n\ n!}{}_1F_1(-n;\gamma;x^2)&=
\psi(\gamma)-\log x^2-(m+1)\sum\limits_{k=0}^m {(-m)_k\over (k+1)^2}\bigg(-{1\over x^2}\bigg)^k\bigg[L_k^{\gamma-1-k}(x^2)
\cr
&-{(\gamma-1)\over x^2}L_k^{\gamma-2-k}(x^2)\bigg]\cr}\eqno(4.3)$$
for ${\alpha\over 2} = 2+m,\quad m =0,1,2,\dots$.
\medskip
\noindent{\bf V. Concluding Remarks}

It is important to notice for $x=0$, the infinite series Eq.(1.13) for $\alpha\geq 2$ indeed diverges. This follows from the fact that ${}_1F_1(-n;\gamma;0)=1$ and 
$$\sum\limits_{n=1}^\infty {({\alpha\over 2})_n\over n}{1\over n!}={\alpha\over 2} \sum\limits_{n=0}^\infty {(1+{\alpha\over 2})_n(1)_n\over (2)_n}{(1)_n\over (2)_n\ n!}={\alpha\over 2}{}_3F_2(1+{\alpha\over 2},1,1;2,2;1)$$
which is absolutely convergent for $\alpha<2$. Therefore, for our results concerning $\alpha=2,4,\dots$ and for the integral representation Eq.(2.9) in general, we must consider $x>0$. The divergence of the infinite series in the expression of the first order perturbation correction of the wavefunction Eq.(1.12) as $x\rightarrow 0$ is indeed controlled by the coefficient term $x^{\gamma-1/2}$ as well by the coefficient $e^{-{x^2\over 2}}$ for $x\rightarrow\infty$. To illustrate the point further, we consider the case of $\alpha=2$, in this case the infinite series in Eq.(1.12) is summable by means of Lemma 1 and the the first-order perturbation correction now reads
$$\psi_0^{(1)}(x)=
{1\over \sqrt{2}}
{1\over (\gamma-1)\sqrt{\Gamma(\gamma)}}
x^{\gamma-{1\over 2}}e^{-{x^2\over 2}}[\log x-{1\over 2}\psi(\gamma)]\quad\quad\quad \gamma>1.\eqno(5.1)
$$
Since $\lim\limits_{x\rightarrow 0}x^{\gamma-{1\over 2}}\log x=0$ for $\gamma>1$, we have $\psi_0^{(1)}(0)=0$. Consequently, the closed form sums of the infinite series Eq.(1.13) contribute for intermediate values $0<x<\infty$ of the wave function rather than the boundaries.

The question posed by Aguillera-Navarro and Guardiola ${\sref{\agub}}$ concerning a special summation formula for Eq.(1.2) in the case of $\alpha =2$ can now be answered with the aid of Lemma 1, which leads to $$\sum\limits_{n=1}^\infty {1\over n}{}_1F_1(-n;{3\over 2};x^2)=\psi({3\over 2})-\log x^2,\eqno(5.2)$$
and the first order perturbation correction is given by means of Eq.(5.1) as
$$\psi_0^{(1)}(x)=
2\pi^{-1/4}xe^{-{x^2\over 2}}[\log x-{1\over 2}\psi({3\over 2})]\eqno(5.3)
$$
which matches the first-order perturbation correction expansion, in powers of $\lambda$, of the exact wave function $\psi_0(x)$, that is Eq.(1.6).

The condition $\gamma>{\alpha\over 2},\quad \alpha=2,4,6\dots,$ imposed on the closed form sums is too strong, for they are indeed valid for weaker conditions. For example,
$$
\sum\limits_{n=1}^\infty{1\over n}{}_1F_1(-n,\gamma,x^2)=
\psi(\gamma)-\log x^2,\ \hbox{
valid for all } \gamma>0,
$$
$$\sum\limits_{n=1}^\infty{n+1\over n}{}_1F_1(-n,\gamma,x^2)=
{\gamma-1\over x^2}-1+\psi(\gamma)-\log{x^2},\ \hbox{
valid for all } \gamma>1,
$$
$$\sum\limits_{n=1}^\infty{{1\over 2} (n^2+3n+2)\over n}{}_1F_1(-n,\gamma,x^2)=
-{3\over 2}+{\gamma-1\over x^2}+{(\gamma-1)(\gamma-2)\over 2x^4}+\psi(\gamma)-\log x^2,\ \hbox{
valid for all } \gamma>2.
$$ 
However the condition has been imposed in order to meet the matrix elements' convergence requirements.

\bigskip
\noindent{\bf Acknowledgment}
\medskip Partial financial support of this work under Grant No. GP3438
from the 
Natural Sciences and Engineering Research Council of Canada is gratefully 
acknowledged by one of us [RLH].

\vfil\eject
\references{1}

\end